\documentclass[showpacs,twocolumn,amsmath,amssymb,prb]{revtex4-1}

\usepackage{graphicx}
\usepackage{amssymb}
\usepackage{amsmath}
\usepackage{color}
\usepackage{color,hyperref}

\newcommand{\bld}[1]{\boldsymbol #1}
\newcommand{\ket}[1]{| #1 \rangle}
\newcommand{\bra}[1]{\langle #1 |}
\newcommand{\rb}[1]{\left( #1 \right)}

\newcommand{\ew}[1]{\langle #1 \rangle}
\newcommand{\beq}{\begin{eqnarray}}
\newcommand{\eeq}{\end{eqnarray}}

\newcommand{\svec}{\mbox{\boldmath$\sigma$}}
\newcommand{\op}[2]{| #1 \rangle \langle #2 |}

\newcommand{\eq}[1]{Eq.~(\ref{#1})}

\begin{document}

\title{Feedback stabilization of pure states in quantum transport}
\author{Christina P\"oltl}
\email{christina@itp.tu-berlin.de}
\author{Clive Emary} 
\author{Tobias Brandes}
\affiliation{
  Institut f\"ur Theoretische Physik,
  Hardenbergstr. 36,
  TU Berlin,
  D-10623 Berlin,
  Germany
}

\date{\today}

\begin{abstract}
We propose a feedback control scheme for generating and stabilizing pure states of transport devices, such as charge qubits, under non-equilibrium  conditions. 
The purification of the device state is conditioned on single electron jumps and 
leaves a clear signal in the full counting statistics which can be used to optimize control parameters. As an example of our control scheme, we are presenting the stabilization pure transport states in a double quantum dot setup with the inclusion of phonon dephasing. 
\end{abstract}
\maketitle

\section{Introduction}
The use of feedback to stabilize desirable structures or behaviors that would otherwise be unobservable is a topic of considerable interest in classical dynamics \cite{KCon1}.
Feedback concepts have been extended into the quantum realm \cite{QCon1,QCon2}, and have been shown to be effective in e.g.  prolonging the life of Schr\"odinger cat states \cite{HK97}, increasing the rate of state reduction under measurement \cite{QCon4a}, and for state purification \cite{QCon4b}. Feedback schemes have also been proposed in the solid state, both theoretically \cite{QCon4c} and in a recent experiment \cite{QCon4d}. 
In these latter examples, a quantum transport device (e.g. quantum point contact, single-electron transistor) acts as a detector of the system (typically a qubit) which is to be controlled. By feeding information gained from the measurement device into the system in a control loop, one can purify the quantum state of the qubit.

\begin{figure}[top]
\includegraphics[width=0.48\textwidth]{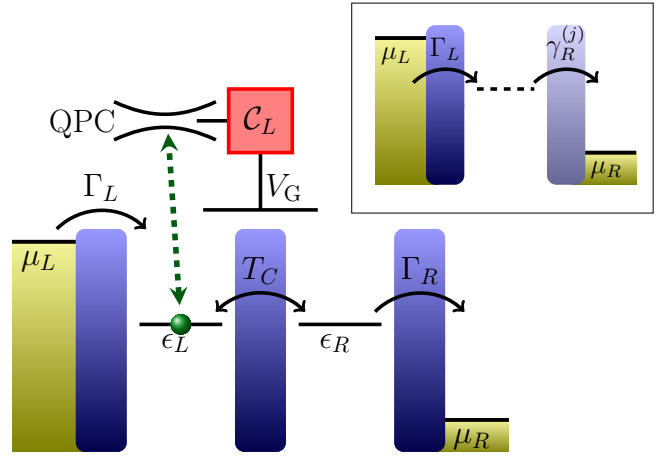}
\caption{\label{Fig1} Main part: After an electron jumps  into the double quantum dot (DQD) (measured e.g. with a quantum point contact QPC) a pulsed control operation modifies e.g. the gate voltages $V_G$ and rotates the electron into a different state. In an effective model this control operation modifies the left jump operator.  
The DQD is described by $\epsilon\mathord=\epsilon_L\mathord-\epsilon_R$ the detuning and  $T_C$ the coupling strength between the dots. The electrons tunnel into/out the left/right dot with the rate $\Gamma_{L/R}$.  
Inset: For sufficient control operations the DQD is stabilized. The transport can then be described by a single resonate level model with an effective right tunnel rate $\gamma_R^{(j)}$.    
  } 
\end{figure}

In this work, we take a different approach and couple information gathered about the charge transport {\em through} a device back into the device itself. 
More specifically, we monitor the flow of electrons from the source lead to a device in the Coulomb blockade regime, as in full counting statistics (FCS)\cite{FCS_exp,FCS1, FCS2, FCS3}, and conditioned on electron jumps, perform unitary operations on the device.  Using this scheme, we show how such feedback can lead to a purification of the device wave function, despite it being in non-equilibrium (this contrasts with the situation without control, where the state of the system is decidedly mixed, as one might expect).  We view this behavior as a genuine state stabilization as these pure states are implicit in the behavior of the device without feedback; but only with the feedback do they become stable.  
Furthermore, we show how successful purification of the transport state also leads to a distinctive signature in the statistics of electron flow through the device.  We describe how this information may be used in the feedback loop to optimize the stabilization.
After presenting a general theory of state stabilization of non-equilibrium quantum transport, we consider the example of a double quantum dot (DQD) (Fig.~\ref{Fig1}) and show that the stabilization of coherent delocalized states is possible. 
%

\section{General theory of the control scheme}
\subsection{Transport model}
We start by considering a general transport Hamiltonian,
$H=H_\mathrm{S}+H_\mathrm{L}+H_\mathrm{T}$,
composed of 
system (e.g. quantum dots (QD)s), leads and tunnel-coupling parts.  We consider two leads with non-interacting Hamiltonian
$
H_\mathrm{L} = \sum_{k,\alpha}\epsilon_{\alpha k} c^\dagger_{\alpha k} c_{\alpha k}
$
with $c_{\alpha k}$ the annihilation operator of an electron in lead $\alpha=L,R$ (left, right) with quantum numbers $k$.
We assume strong Coulomb blockade and a bias configuration such that at most one excess electron can be in the system at any given time.  The state-space of the system is thus spanned by the empty state $\ket{0}$ and $N$ single electron states $\ket{n}$, typically not eigenstates of system Hamiltonian $H_\mathrm{S}$.  Finally we assume that the leads are each coupled to one and only one localized system level ($\ket{L}$ and $\ket{R}$ respectively) such that the tunnel Hamiltonian reads
$
  H_\mathrm{T} = \sum_{ k} 
  t_{R k} c_{R k}^\dagger D_{R}
  +t_{L k} c_{L k}^\dagger D_{L}^\dag
  +\mathrm{H.c.}
$
with  $D_{L}^\dagger\mathord=\op{L}{0}$ and $D_{R}^\dagger \mathord=\op{0}{R}$.

In the high-bias limit with tunnel rates $\Gamma_{\alpha} =2\pi  \sum_{k}  |t_{\alpha k}|^2 \delta(\epsilon-\epsilon_{\alpha k})$ assumed constant, the behavior of the system density matrix can be described with a Markovian master equation of Lindblad form 
 \cite{Gur}:
\beq\label {master0}
  \dot\rho = {\cal W}\rho = \Big( {\cal W}_0 
  +
  {\cal J}_{L} 
  + 
  {\cal J}_{R} \Big) \rho
  , 
\eeq
with jump super-operators 
$
  {\cal J}_{\alpha}\rho\mathord
  = \Gamma_\alpha D^\dag_{\alpha} \rho D_{\alpha}
$, which transfer an electron to/from lead $\alpha = R/L$, and free Liouvillian ${\cal W}_0$, which describes the evolution of the system without electron transfer.  This latter assumes the form
 $ {\cal W}_0 \rho = 
  -i
  \left\{
    \widetilde{H} \rho - \rho \widetilde{H}^\dag
  \right\}
  \label{free-ev}
 $ (we set $\hbar=1$),
where
\beq
  \widetilde{H} = H_S 
  - i \frac{1}{2} \sum_\alpha \Gamma_\alpha  D^\dag_\alpha D_\alpha 
, \label{newH}
\eeq
is an effective {\it non-Hermitian} `Hamiltonian' operator for the system.

\subsection{Electron counting and control }
The evolution of the density matrix under \eq{master0} can be interpreted  in terms of trajectories\cite{qu-traj}. The density matrix at time $t$ is then given as a sum of terms such as
\begin{align}
\int^t_0 dt_2 
  \int^{t_2}_0 dt_1 
  \Omega_0(t-t_2) {\cal J}_{R} \Omega_0(t_2-t_1) {\cal J}_{L}\Omega_0(t_1) \rho(t_0), \label{traj}
\end{align}
which describe all trajectories in which an electron jumps into the system at time $t_1$ ($ {\cal J}_{L}$), out again at time $t_2$ ($ {\cal J}_{R}$), with evolution in between described by the no-jump propagator $\Omega_0(t) = e^{{\cal W}_0 t}$.
As in FCS experiments, we assume that our measurement device is sensitive to single electron jumps.  Conditioned then on the detection of an electron jump from the left lead, we set the control electronics to implement a control operation on the system which we consider here to act as an instantaneous unitary operation on the system \cite{QCon1}. In an experimental setup for the transport through QDs these control operations could be achieved, for example, with pulsed gate voltages.
This scheme modifies the quantum trajectories such that each left jump operator is followed by a control operation; we thus replace
${\cal J}_{L}$ with ${\cal J}_{L}^C={\cal C}_{L}  {\cal J}_{L}$ in all trajectories like \eq{traj}.  Re-summing all the trajectories, we arrive at an effective Liouvillian for the evolution of the density matrix with control that is the same as ${\cal W}$ in \eq{master0} but with ${\cal J}_{L} $ replaced by  ${\cal J}_{L}^C$.


\subsection{Feedback Stabilization}  

We now describe how this control schema can be used to stabilize certain states.
The effective ``Hamiltonian'', \eq{newH}, has right and left eigenstates 
$   \widetilde{H} \ket{\psi_j}= \varepsilon_j \ket{\psi_j} $ and 
$    \bra{\widetilde{\psi_j}} \widetilde{H}= \varepsilon_j \bra{\widetilde{\psi_j}}$, 
which, in general, are non-adjoint: $\bra{\widetilde{\psi_j}} \ne \rb{\ket{\psi_j}}^\dag$,
and can be used to write $\widetilde{H}$ in its diagonal basis as
\beq
   \widetilde{H}  
   =  
   - i \frac{1}{2} \Gamma_{L} \op{0}{\widetilde{0}}
   +\sum_{j=1}^N\varepsilon_j \op{\psi_j}{\widetilde{\psi_j}}
,
\eeq
where we have separated the empty state since it is not coupled to other states by $\widetilde{H}$. 
The eigen-operators of the free Liouvillian are
$\rho_{jk}\mathord = \op{\psi_j}{\psi_k}$ such that
${\cal W}_0 \rho_{jk}\mathord = -i \rb{\varepsilon_j - \varepsilon_k^*}\rho_{jk}$.
The diagonal matrices $\rho_{jj}$ represent pure states and obey 
$
  {\cal W}_0 \rho_{jj}\mathord =2 \Im( \varepsilon_j) \rho_{jj}
$.
The empty state is one such eigen-operator with
$
  {\cal W}_0\rho_{00} = -\Gamma_L \rho_{00} 
$.
The remaining pure density matrices $\rho_{jj};~j>0$ are states that we will seek to stabilize with our feedback scheme.

To effect the state stabilization, we choose the control operation ${\cal C}_{L}$ to rotate   state $\ket{L}$ into the desired eigenstate of $\widetilde{H}$.
The effective left jump operator therefore acts as
$
  {\cal J}^C_{L} \rho_{00} = \Gamma_L \rho_{jj},
$
where the $\Gamma_L$ forefactor arises from the normalization and the fact that the empty state decays with rate $\Gamma_L$ with or without control.
Once in state $\rho_{jj}$, no transitions out of this state are induced by ${\cal W}_0$.  Rather, the only thing that happens to an electron in this state is that it leaves at a rate
 $\gamma_R^{(j)} = - 2 \Im(\varepsilon_j)$.
By jumping directly into one of the eigenstates of the free Liouvillian $\ket{\psi_j}$ the dynamic of the system is then determined solely by the vacuum $\ket{\psi_0}$ and $\ket{\psi_j}$; the other states are decoupled. In the basis $\{\rho_{00},\rho_{jj}\}$ the Liouvillian of the feedback controlled system  in the above parametrization  can be written as
\beq \label{Leff}
  {\cal W}^{(j)}_C = 
  \rb{
  \begin{array}{cc}
    -\Gamma_{L} & \gamma_R^{(j)} \\
    \Gamma_L & - \gamma_R^{(j)}
  \end{array}
  }.
\eeq
This Liouvillian corresponds to an effective single resonant level, with the effective tunnel rate $\gamma_R^{(j)}$ and the stationary state
\beq
  \rho_\mathrm{stat} = 
  \frac{1}{\Gamma_{L} + \gamma_R^{(j)}}
  \rb{
  \gamma_R^{(j)}\op{0}{0} 
    + 
    \Gamma_{L}\op{\psi_j}{\psi_j}
  }.
\eeq
In the limit $\Gamma_{L} \gg \gamma_R^{(j)}$ the stationary state reduces to
\beq
   \lim_{\Gamma_{L}/\gamma_R^{(j)} \to \infty}\rho_\mathrm{stat} =\op{\psi_j}{\psi_j},
\eeq
and the system is thus {\em stabilized} in the pure state $\ket{\psi_j}$.  

In general then, a system with $N$ internal states has $N$ pure states that can be prepared in this manner. These states depend on the internal system parameters as well as on $\Gamma_R$.  In the limit of $\Gamma_R\mathord\rightarrow 0$, the stabilized states will be eigenstates of $H_\mathrm{S}$.  For finite $\Gamma_R$, however, these states are, in some sense, non-equilibrium pure states, since they depend on the rate $\Gamma_R$ and can be quite different from the eigenstates of the bare Hamiltonian $H_\mathrm{S}$.  Needless to say, the states are unstable without control.


\section{Feedback stabilization of a double quantum dot} 
In the following, we discuss the $N\mathord=2$ example of a charge qubit with Hamiltonian    
$
 H_\mathrm{S}=\frac{1}{2}\epsilon \sigma_z +T_C\sigma_x , 
$
with $\sigma_z \mathord=\op{L}{L}\mathord-\op{R}{R}$, $\sigma_x\mathord= \op{L}{R}\mathord+\op{R}{L}$, $\epsilon\mathord=\epsilon_L\mathord-\epsilon_R$ the detuning and  $T_C$ the coupling strength between the dots. 
In the infinite bias limit \cite{Gur,DQD1,Bra05} the transport through the DQD is described by \eq{master0}
with $D_L^\dagger \mathord= \op{L}{0}$ and $D_R^\dagger \mathord= \op{0}{R}$. The corresponding Liouvillian of the DQD is shown in appendix~\ref{App1}.

\subsection{Zero detuning $\epsilon=0$} 
First we consider the case without detuning $\epsilon\mathord=0$ for simplicity.
The right-eigenvalues of $\widetilde{H}$ are $\varepsilon_0\mathord=\mathord-i \frac{\Gamma_L}{2}$ 
with eigenvector $\ket{\psi_0}\mathord=\ket{0}$ and 
\beq \label{eigendot}
\varepsilon_{\mp}=\frac{1}{4}\big(-i\Gamma_R\mp \kappa   \big),\quad \kappa\mathord=\sqrt{16T_C^2-\Gamma_R^2},
\eeq
with the eigenvectors
$\ket{\psi_\mp}=\mathcal N_{\mp}^{-1}\big[ (i\Gamma_R\mp \kappa)  \ket{L}+4T_C\ket{R}\big]$,
normalized with 
$\mathcal N_{\mp}\mathord=\sqrt{16T_C^2+|i\Gamma_R \mp \kappa |^2}$. 

In the following, we consider the simplification $\Gamma_L \gg  \Gamma_R,T_C $ which allows us to effectively project out the empty 
dot state $\ket{\psi_0}$ and to assess the purity of the transport qubit in the usual Bloch representation  
with $\ew{\sigma_x}\mathord= 2\Re(\rho_{LR})$,
$\ew{\sigma_y}\mathord= 2\Im(\rho_{LR})$ and $\ew{\sigma_z}\mathord= \rho_{LL}\mathord-\rho_{RR}$.

\begin{figure}[top]
\includegraphics[width=0.48\textwidth]{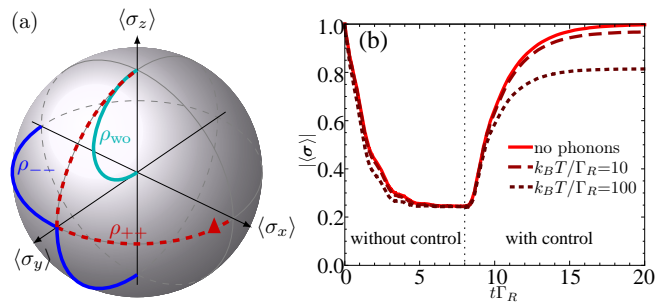}
\caption{
(a) Stabilisable double quantum dot states $\rho_{\mp\mp}$   on the surface of the Bloch sphere, and stationary state $\rho_{\rm wo}$ without control, for tunnel coupling $T_C\in \{0,\infty\}$ . The triangle marks the state which is stabilized in (b). 
(b) 
Time evolution of  Bloch vector length $|\ew{\svec}|$ with control operation switched on at $t=8 \Gamma_R^{-1}$ for different temperatures $T$. Parameters: $T_C=2\Gamma_R$, phonon coupling  $g=8\cdotp 10^{-4}$, Debye-cutoff $\omega_c=500\Gamma_R$ and control parameters $\theta=\arccos(3 \sqrt{7/127})$ and 
$\theta_C=\text{arcsec}(8 \sqrt{2})$.
 \label{Fig2}} 
\end{figure}

 The pure-state density matrices $\rho_{\mp\mp}\mathord=\op{\psi_\mp}{\psi_\mp}$ to be stabilized correspond to two of the occupied eigenstates of the free Liouvillian  ${\cal W}_0$.
 Fig.~\ref{Fig2}(a)  depicts these stabilisable pure states on the surface of the Bloch sphere.
For $\Gamma_R\mathord>4 T_C$, they describe  partly localized states 
with 
\begin{align}
 \ew{\sigma_z}\mathord=&  \mp \frac{\tilde{\kappa}}{\Gamma_R},& 
\ew{\sigma_x}\mathord=&0\quad &&\text{and} & 
\ew{\sigma_y}\mathord=&\frac{4T_C}{\Gamma_R}, \label{smallTc}
\intertext{and for $\Gamma_R\mathord<4 T_C$
completely delocalized states with}
\ew{\sigma_z}\mathord=&  0,& 
\ew{\sigma_x}\mathord=&\mathord \mp \frac{\kappa }{4T_C} &&\text{and}& 
 \ew{\sigma_y}\mathord=&\frac{\Gamma_R}{4T_C}, \label{bigTc} 
\end{align}
with $\tilde{\kappa}\mathord=\sqrt{\Gamma_R^2-16T_C^2}$.	
At $\Gamma_R\mathord=4T_C$ the eigenvalues $\varepsilon_\mp$ are degenerate and the corresponding eigenstates cross each other  with $\ew{\sigma_z}\mathord=0$, $\ew{\sigma_x}\mathord=0$ and $\ew{\sigma_y}\mathord=1$.

\begin{figure}[top]
\includegraphics[width=0.48\textwidth]{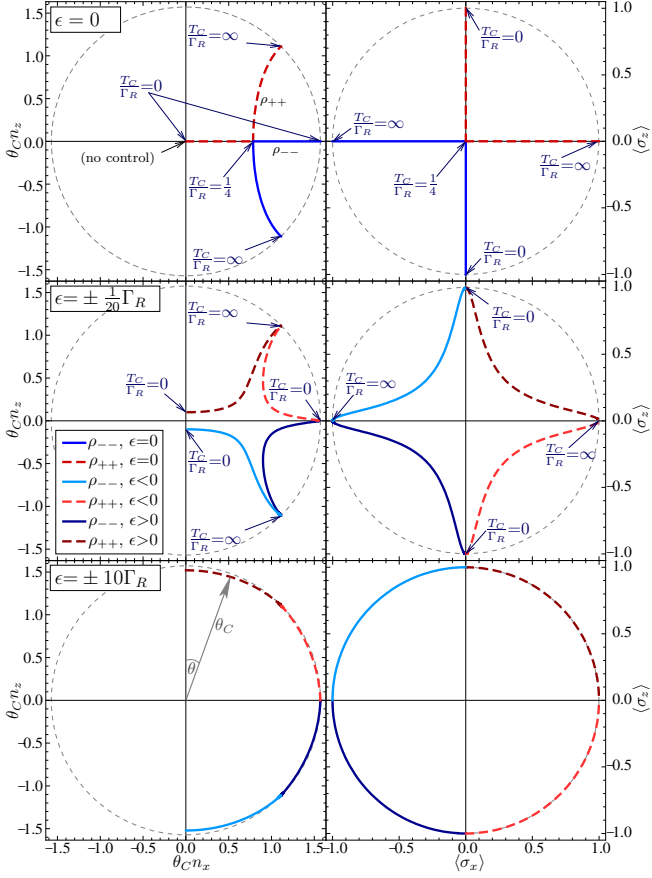}
\caption{\label{Fig3n} (left) 
The control operations needed to stabilize states at different detunings $\epsilon$ for $T_C\in \{0,\infty \}$ are shown on the control plane. The control strength $\theta_C$ is determined by the length of a vector between a point on the control plane and the origin. The angle between the positive part of the $\theta_C n_z$-axis and the vector is the control angle $\theta$.   
(right) The corresponding stabilizable states  
are shown as projection on the $\ew{\sigma_z}$-$\ew{\sigma_x}$- plane of the Bloch sphere.   } 
\end{figure}

Without control an electron simply tunnels into the left dot (and thus into a localized, pure state). 
The goal of the control operation is to rotate this state into one of the pure states $\rho_{\pm\pm}\mathord \equiv  \op{\psi_\pm}{\psi_\pm}$. 
In general, we can consider every qubit rotation as a possible control operation ${\cal C}_L \mathord=e^{\theta_C \bld{n}_0\cdot \vec{\Sigma}}$ where $\vec{\Sigma}$ corresponds to the Pauli matrices in Liouville space $\vec{\Sigma}\rho\mathord=\mathord-i2[ \svec,\rho]$ with the usual Pauli matrix vector $\svec$,
the unit vector $\bld{n}_0\mathord=\{n_x,n_y,n_z\}\mathord=\{\sin{\theta},0,\cos{\theta}\}$  that determines the angle of the rotation in the $\sigma_x$-$\sigma_z$ plane, and the control strength parameter $\theta_C$. 
 A straightforward calculation (see appendix \ref{App2}) leads to the control parameters $\theta\mathord=\pi/2$ and $\theta_{C}\mathord=\arccos\rb{\sqrt{{(\Gamma_R \pm \tilde{\kappa})}/{2\Gamma_R}}}$ for $\Gamma_R\mathord>4 T_C$, and 
$\theta\mathord=\arccos\rb{\mathord\pm\tilde{\kappa}/\sqrt{{\Gamma_R^2-32T_C^2}}}$ and  
$\theta_{C}\mathord=\text{arcsec}\rb{{4\sqrt{2}T_C}/{\Gamma_R}}$ for $\Gamma_R\mathord<4 T_C$, corresponding to the stabilization of $\rho_{\pm\pm}$. The upper left plot of Fig.~\ref{Fig3n} shows these control parameter on the control plane.

The fidelity of our control scheme is characterized by the length of the Bloch vector $|\ew{\svec}|$, obtained by solving \eq{master0} for the DQD model. 
Fig.~\ref{Fig2}(b)(solid line) demonstrates the effect of the control operation $\mathcal{C}_L$ on the DQD Bloch vector length $|\ew{\svec}|$ --- switching on the control (at time $t=8\Gamma_R^{-1}$ in the plot) leads to a rapid purification of the transport qubit state, even though electrons continuously tunnel through the DQD.

\subsection{Finite detuning $\epsilon\neq 0$}
With finite detuning it is possible to stabilize all states on the Bloch sphere which have a positive $\ew{\sigma_y}$ component. The exact expressions for the stabilizable  eigenstates with $\epsilon\neq0$ can be found in appendix \ref{App2}. 

Fig.~\ref{Fig3n} (right) shows the projection of the stabilizable eigenstates on the $\ew{\sigma_z}$-$\ew{\sigma_x}$- plane of the Bloch sphere for different detunings for $T_C\in\{0,\infty\}$. Without detuning ($\epsilon=0$) the stabilizable states lie on the axis of the plane which can be also seen in Fig.~\ref{Fig2}(a). 
Since all stabilized states have a Bloch vector with length one and a positive $\ew{\sigma_y}$- component, the size of $\ew{\sigma_y}$ is maximal when the states in the $\ew{\sigma_z}$-$\ew{\sigma_x}$- plane have the shortest distance to the origin. 
For increasing detuning these states move asymptotically towards the unit circle, meaning the maximal size of the $\ew{\sigma_y}$ component decreases for increasing $|\epsilon|$. Each of the four quadrants of the plane is covered by a different  solution: The I- quadrant by the $\rho_{++}$ solution for $\epsilon>0$, the II- quadrant by the $\rho_{--}$ solution for $\epsilon<0$, the III- quadrant by the $\rho_{--}$ solution for $\epsilon>0$ and the IV- quadrant by the $\rho_{++}$ solution for $\epsilon<0$.   
The control operation necessary to stabilize the states for each detuning are shown on the left side of Fig.~\ref{Fig3n}.

\subsection{Phonons}
In any realistic experimental setup, our ideal coherent feedback control scheme will be disturbed by the coupling to a dissipative environment. In order to make quantitative predictions, we therefore introduce an additional coupling of the DQD  to a bath of thermal phonons \cite{DQD3}  with wave vector $Q$ and energy $\omega_Q$. We describe the electron-phonon coupling  
via a (pseudo) spin-boson Hamiltonian \cite{Bra05} 
\begin{align}
H_{\text{phon}} =& \sum_{Q}\Big[ \frac{1}{2} \sigma_z g_Q(a_{-Q}+a^\dagger_Q)+\omega_Q a^\dagger_Q a_Q \Big]
\end{align}
that leads to a modification of the free Liouvillian ${\cal W}_0$ in our master equation \eq{master0}. These modifications are included in the free Liouvillian shown in appendix~\ref{App1}. We assume that an electron in the DQD couples to a phonon bath with an ohmic spectral density with exponential cut-off at Debye frequency  $\omega_c$.
Without detuning ($\epsilon=0$) the corresponding dephasing rates at temperature $T$ (we set the Boltzmann constant $k_B=1$) are
\beq\label{gammap}
\gamma_p = \gamma \coth \frac{T_C}{T},\quad \gamma \equiv  g \pi T_C e^{\frac{-2T_C}{\omega_c}}, \quad \gamma_b=0,
\eeq
where $g\ll 1$ is a dimensionless  coupling parameter.

Fig.~\ref{Fig2}(b) also shows the evolution of the fidelity with the inclusion of the phonon bath (dashed and dotted line) at finite temperature. The lengths of the stabilized steady-state Bloch-vectors with phonons is below one $|\ew{\bld{\sigma}}|<1$ since the phonon dephasing limits the stabilization of the pure states. 

%
\begin{figure}[top]
\includegraphics[width=0.48\textwidth]{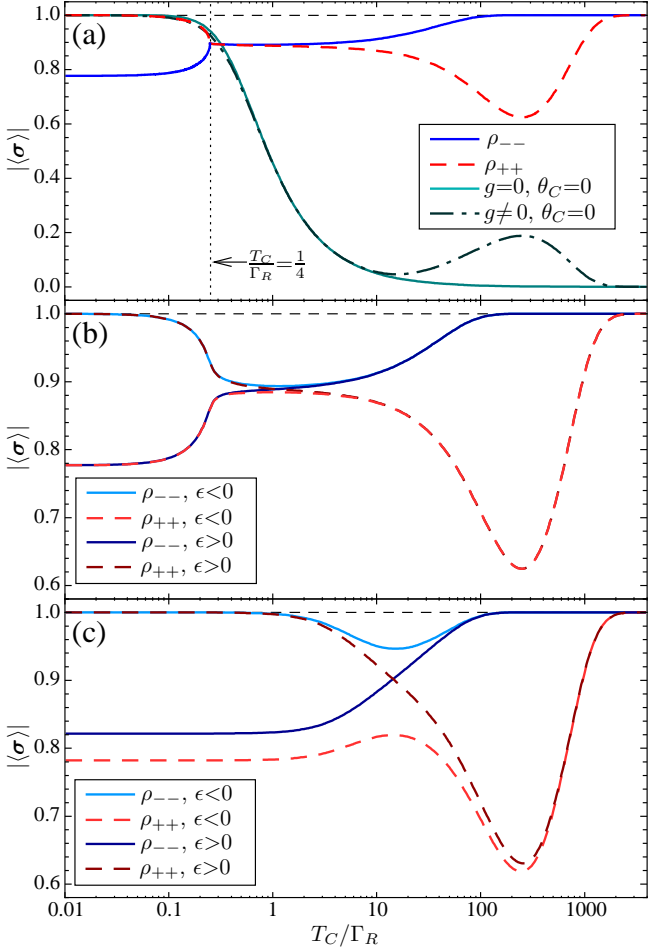}
\caption{\label{Fig4} Loss of stabilization of the  control stabilized pure states $\rho_{\mp\mp}$ due to phonons illustrated by the length of the Bloch vector as a function of $T_C$ for different detunings (a) $\epsilon=0$ (without detuning), (b) $\epsilon=\pm\frac{1}{20}\Gamma_R$ and (c) $\epsilon=\pm10\Gamma_R$. Without control ($\theta_C=0$) the length of the Bloch vector decays to zero for large tunnel couplings $T_C$.  
Parameters: $\Gamma_L\rightarrow \infty$, phonon coupling $g=4\cdotp 10^{-4}$, $\omega_c/\Gamma_R=500$, $k_B T=50\Gamma_R$.  } 
\end{figure}

Fig.~\ref{Fig4}(a) shows the fidelity of the steady state for $\epsilon=0$
as a function of tunnel coupling $T_C$.  The stationary state without control ($\theta_C\mathord=0$) is always mixed (except for trivially at $T_C=0$ where $|\ew{\svec}|\mathord =  1$ as then an electron tunnels into the left dot and never leaves it), with $|\ew{\svec}|\mathord\to 0$ at very large $T_C$. 
In contrast the control-stabilized states $\rho_{\pm\pm}$ remain pure in absence of phonons  ($|\ew{\svec}|\mathord=1$ for $g=0$). 
These pure states  are differently affected by the phonon dephasing.
This striking behaviour of  the two phonon branches for the  control-stabilized states $\rho_{\mp\mp}$ in Fig.~\ref{Fig4}(a) can be traced back to the character of the eigenvalues $\varepsilon_{\mp}$, \eq{eigendot},
and the corresponding eigenstates $\ket{\psi_\mp}$ of $\widetilde{H}$. Starting from large  $T_C\gg \Gamma_R/4$, the $\varepsilon_{\mp}$ have a much larger real than imaginary part and $\ket{\psi_-}$ ($\ket{\psi_+}$)   essentially corresponds to the ground (excited) state of the isolated DQD.  In this regime of large level splitting, spontaneous emission of phonons from $\ket{\psi_+}$ to $\ket{\psi_-}$ dominates phonon absorption, 
with the purity of  $\rho_{++}$ strongly reduced as compared with the one of  $\rho_{--}$, which is shown by the explicit expressions for the fidelities $|\ew{\svec_{\pm}}|$ of the two states,
\begin{align}
|\ew{\svec_{\pm}}|\approx1-\rb{\frac{2\gamma_p}{\Gamma_R}-\frac{4\gamma_p^2}{\Gamma_R^2}}
\left(1\pm \tanh\!\frac{T_C}{T}\right). \label{appTcbig}
\end{align}
In particular, the shape of the $\rho_{++}$ branch for large $T_C$ and at low temperatures is entirely determined by the spectral density of the phonons $\propto T_C e^{-2T_C/\omega_c}$ and can therefore be used 
as a spectroscopic tool to determine the dephasing rate $\gamma_p$, \eq{gammap}. 

Moving to smaller tunnel coupling $T_C$,  the  two broadened levels of the DQD start to overlap at $T_C \mathord= \Gamma_R/4$ at which point the $\varepsilon_{\mp}$, \eq{eigendot},  become purely imaginary, and the two merged branches split again at still smaller $T_C$. In absence of phonons, our control scheme coherently rotates the DQD into the pure states $\ket{\psi_\pm}$, i.e.,  
$|\ew{\svec_{\pm}}|\mathord=1$ for $g\mathord=0$. The presence of phonons ($g\mathord> 0$), however, leads to an un-controlled decay of $\ket{\psi_\pm}$ that can be 
described by  $\Gamma_R\to  \Gamma_R\mathord+2\gamma_p$ in the imaginary eigenvalues \eq{eigendot}, which for $T_C \mathord\ll \Gamma_R$ then become $\varepsilon_{+} \mathord\approx  0 $ (very slow decay) and $\varepsilon_{-}\mathord \approx  -i(\Gamma_R/2+\gamma_p) $ (fast decay). This `repulsion' of imaginary levels is analogous to the Dicke spectral line effect \cite{Bra05,Dic53}, i.e. a splitting into a sub- and a superradiant decay channel, and is again confirmed by the fidelities  
\beq
|\ew{\svec_{+}}|\approx 1,\quad
|\ew{\svec_{-}}|\approx 1-\frac{4\gamma_p}{\Gamma_R}-\frac{8\gamma_p^2}{\Gamma_R^2},
\eeq
valid 
for $T_C \mathord\ll \Gamma_R$.

The fidelity with finite detuning as a function of $T_C$ is shown in Fig.~\ref{Fig4}(b) and (c). The main features of the plots are the same as in the case without detuning Fig.~\ref{Fig4}(a). However if the detuning between the dots is positive $\epsilon>0$ the $\rho_{++}$- and the $\rho_{--}$- branch cross each other, while for negative detunings $\epsilon<0$ they preform an anti-crossing. In general the stabilizable states (see Fig.~\ref{Fig3n}(right)) for small couplings $T_C\rightarrow0$ lie  close to $\ew{\sigma_z}=\pm 1$ and for large $T_C\rightarrow\infty$ close to $\ew{\sigma_x}=\pm1$. The phonon dephasing has the strongest effect on the states which either lie close to $\ew{\sigma_z}=-1$ or on the states with $\ew{\sigma_x}\lesssim1$.
The control operation in the first case rotates the electron that originally tunnels in the left dot into a state which is almost entirely localized in the right dot and in the second case the electron is rotated into a state which corresponds to the excited state of the isolated system (see discussion for $\epsilon=0$). In both cases the states are strongly effected by dephasing. In the first case this effect vanishes for decreasing temperatures but in the second case spontaneous emission even at zero temperature destabilizes the states, with the exception of states above the Debye cut-off (these states are only weakly effected by the phonon dephasing). This knowledge explains the anti-crossing  and crossing in Fig.~\ref{Fig4}(b) and (c). For $\epsilon<0$ the solution stronger destabilized by phonons is the same ($\rho_{++}$) for small and for large $T_C$ which leads to an anti crossing and for $\epsilon>0$ the solution stronger affected by dephasing switches form $\rho_{--}$ for small $T_C$ to $\rho_{++}$ for large $T_C$ resulting in a crossing. 

\begin{figure}[top]
\includegraphics[width=0.5\textwidth]{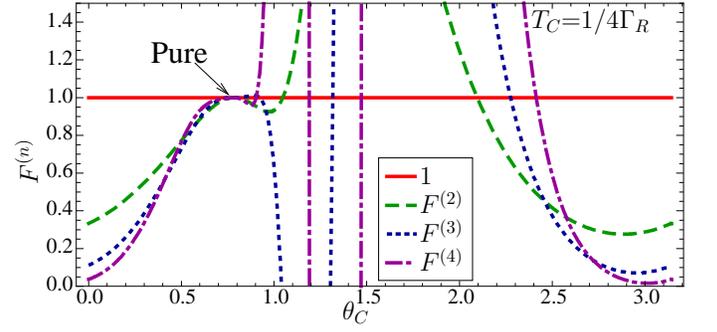}
\caption{Fano factors $F^{(n)}$ for the double quantum dot with zero phonon coupling 
as a function of the control strength $\theta_C$ at fixed qubit rotation angle $\theta=\frac{\pi}{2}$. 
Stabilization of a pure state     
occurs at the value of $\theta_C$ where $F^{(n)}=1; \forall n>0$, corresponding to Poissonian statistics in the limit of $\Gamma_L\gg T_C,\Gamma_R$ considered here. ($\epsilon=0$)
\label{Fig5}} 
\end{figure}

\subsection{Full counting statistics}  
Collecting all trajectories with $n$ left-jumps as partial density matrix $ \rho^{(n)}$, \eq{master0} implies the number-resolved master equation for the system with control
\beq \label{nres}
  \dot{\rho}^{(n)} =  
 ( {\cal W}_0 + {\cal J}_{R})\rho^{(n)} 
  + {\cal J}^C_{L}\rho^{(n-1)}.
\eeq
With the Fourier transform  $\rho(\chi) = \sum_{n} e^{i n \chi} \rho^{(n)}$ we obtain the $\chi$-resolved master equation
\beq\label {master1}
  \dot\rho(\chi)
  = \Big( {\cal W}_0 + 
  {\cal J}_{R}
  +
   {\cal J}^C_{L} e^{i\chi } 
  \Big) \rho(\chi). 
\eeq
Using the current cumulant generating function\cite{FCS1, FCS2, FCS3} 
\begin{align}
 {\cal F}(\chi,t)= \ln\Big(\text{Tr}_D\big\{ e^{({\cal W}_0 + 
  {\cal J}_{R}
  +
   {\cal J}^C_{L} e^{i\chi })(t-t_0)}\rho(t_0)\big\} \Big),
\end{align}
where $\text{Tr}_D\{\cdots\}$ corresponds to the trace of density matrix, we calculate the $n$-th order zero frequency current correlation
\begin{align}
\ew{S^{(n)}}=\frac{\text{d}}{\text{d}t}\frac{\partial^n}{\partial (i\chi)^n} 
{\cal F}(\chi,t) |_{\chi=0,t\rightarrow\infty}. 
\end{align}
We define the zero-frequency Fano factors as
\begin{align}
 F^{(n)}=\frac{\ew{S^{(n)}}}{\ew{S^{(1)}}},
\end{align}
the $n$-th order current correlation functions, normalized by the stationary current \cite{FCS3} ($\ew{S^{(1)}}=\ew{I}$).  

Fig. \ref{Fig5} shows the zero-frequency Fano factors $F^{(n)}$  with $n=2,3,4$ for our DQD example as a function of control strength $\theta_C$ with fixed static dot parameters and control angle $\theta=\frac{\pi}{2}$ (again in the limit $\Gamma_L \to \infty$).  Whilst the cumulants exhibit a rather complex dependence on $\theta_C$ at exactly the point where the pure qubit state is stabilized, all Fano factors are simply $F^{(n)}=1$.
This can be understood by recalling that at the stabilization point, the system is equivalent to a  single resonant level in its transport properties, cf. \eq{Leff}. The effective outgoing rate for our DQD example for $\epsilon=0$ reads
\beq
  \gamma_R^{(\mp)}&=&
  \left\{
  \begin{array}{cc}
    \frac{1}{2}\Gamma_R & \Gamma_R\mathord<4T_C\\
    \frac{1}{2}(\Gamma_R\pm\sqrt{\Gamma_R^2-16T_C^2} ) & \Gamma_R\mathord>4T_C
  \end{array}
  \right. .
\eeq
The cumulant generating function of this effective model  in the long time limit is well-known \cite{FCS2,FCS3}
\begin{align}
 \mathcal F(\chi,t)=&  \frac{t}{2}\Big(-\mathord\Gamma_L\mathord-\gamma_R^{(j)}\mathord+\sqrt{(\Gamma_L-\gamma_R^{(j)})^2 +4\Gamma_L \gamma_R^{(j)}e^{i\chi}}\Big). \label{SET-FCS}
\end{align}
Which in the limit $\Gamma_L \to \infty$ reduces to the cumulant generating function of a Poissonian process for which all Fano factors are equal to unity.

This information that the stabilized state has such a distinctive FCS can be used to locate the control operation required  for stabilization {\em even in the absence of knowledge of the precise model parameters}.  Knowledge of the electron jumps permits the FCS of the transport to be calculated.  For a given set of, presumably non-stabilizing, control parameters, this FCS can be compared with an ideal Poissonian signal as a target. The control parameters can then be adjusted and the FCS compared again.  This can be implemented in an {\em in situ} (classical) closed feedback loop to optimize the control operation such as to permit stabilization.  
With electron jumps occurring, e.g., on time scales of milli-seconds in state-of-the-art FCS experiments \cite{Flietal09}, no ultra-fast electronics is necessary for information processing.
The optimization should also work if the condition of large  $\Gamma_L$ can not be assumed, in which case the target signal would no longer be Poissonian but given by the cumulants 
of the two coupled Poisson processes (left and right tunnel barrier) of the resonant level model \eq{SET-FCS}, Ref.~\cite{FCS2,FCS3}.


\section{Conclusions} 
We have presented a general theory for the stabilization of pure states in a non-equilibrium electron transport setup.  The stabilized states have the features of a single resonant level including the FCS, which can be used to detect these pure states. 
We demonstrated this stabilization on the example of a DQD and have shown that half of the states on the Bloch sphere can be stabilized.
Moreover the interaction with an dissipative environment due to the coupling to a phonon bath was discussed. 
Further generalization of the control stabilization to systems with many body states and more leads enable to use this stabilization scheme for arbitrary Coulomb blockade systems.

We are grateful to G.~Kie{\ss}lich, G.~Schaller, A.~Beyreuther and 
K.~Mosshammer for useful discussions. Financial support by DFG projects BR 1528/7, BR 1528/8, GRK 1558 and SFB 910 is acknowledged.

 \appendix
 \section{Double quantum dot Liouvillian }\label{App1}
We present the DQD Liouvillian with phonon coupling in the pseudo-spin basis with $\rho=\{\rho_{00},n_{\text{occ}},\ew{\sigma_x},\ew{\sigma_y},\ew{\sigma_z} \}=\{\rho_{0},\rho_{LL}+\rho_{RR},\rho_{LR}+\rho_{RL},\frac{1}{i}(\rho_{LR}-\rho_{RL}),\rho_{LL}-\rho_{RR} \}$: 
\begin{align}
\mathcal W_0 = 
&  
\begin{pmatrix}
-\Gamma_L & 0 & 0& 0 & 0 \\
                          0 & -\frac{1}{2}\Gamma_R & 0 & 0& \frac{1}{2}\Gamma_R \\
                           0 & -\gamma &-\frac{1}{2}\Gamma_R-\gamma_p & \epsilon& \gamma_b \\
                           0 &  0 & -\epsilon &- \frac{1}{2}\Gamma_R-\gamma_p &2T_C \\
                           0 & \frac{1}{2}\Gamma_R & 0 & -2T_C &- \frac{1}{2}\Gamma_R 
\end{pmatrix},   \\
\mathcal J_L^C = 
&  
\begin{pmatrix}
0 & 0 & 0& 0 & 0 \\
                          \Gamma_L & 0 & 0 & 0 &0 \\
                           \Gamma_L \sin(2 \theta) \sin(\theta_C)^2&0 & 0 & 0 & 0 \\
                           \Gamma_L \sin(\theta)\sin(2\theta_C) &  0 & 0 & 0 & 0 \\
                           \Gamma_L\big(\cos(\theta)^2+\cos(2\theta_C)\sin(\theta)^2\big) & 0 & 0 & 0 & 0 
\end{pmatrix},   \\
\text{and} &  \quad \mathcal J_R = 
  \begin{pmatrix}
0 & \frac{1}{2}\Gamma_R & 0& 0 & -\frac{1}{2}\Gamma_R \\
                          0 & 0 & 0 & 0 &0 \\
                           0&0 & 0 & 0 & 0 \\
                           0 &  0 & 0 & 0 & 0 \\
                           0 & 0 & 0 & 0 & 0 
\end{pmatrix}.  
\end{align}
The left jump operator without control ${\cal J}_L$ can be calculated by setting $\theta_C=0$ in ${\cal J}_L^C$. 
For an Ohmic spectral density with exponential cut-off at Debye frequency  $\omega_c$, the phonon dephasing rates are
\begin{align}
\gamma_p =&\frac{g\pi}{\Delta^2}\left[\frac{\epsilon^2}{\beta}+2T_C^2\Delta e^{-\Delta /\omega_c}
\coth{\left(\frac{\beta\Delta}{2}\right)}\right],\nonumber\\
\gamma =&g\pi T_C e^{-\Delta /\omega_c}, \nonumber\\
\gamma_{b} =&g\frac{\pi T_C}{\Delta^2}\left[\frac{2 \epsilon}{\beta}
-\epsilon\Delta e^{-\Delta /\omega_c}\coth\left(\frac{\beta\Delta}{2}\right)\right],
\end{align}
where $g\ll 1$ is a dimensionless  coupling parameter, $\beta=(k_B T)^{-1}$ is the inverse of temperature $T$ times Boltzmann constant $k_B$ and $\Delta = \sqrt{\epsilon^2+4T_C^2}$.
In order to find the stabilizable states the phonon coupling has to be set to zero ($g=0$ or $\gamma=\gamma_p=\gamma_b=0$).

\section{Stabilizable states}\label{App2}
The DQD- eigenstates we seek to stabilize are $\rho_{\pm \pm}$. The control operation which stabilizes these states are determined by $\frac{{\cal J}_L^C}{\Gamma_L} \rho_{00}= \rho_{\pm \pm}$. 
In general, the right eigenvalues of $\widetilde{H}$ of the DQD which correspond to occupied states are   
\begin{align}
 \varepsilon_{\mp}=\frac{1}{4}\big(-i\Gamma_R\mp\sqrt{4\epsilon^2+4i\epsilon\Gamma_R-\Gamma_R^2+16T_C^2}\big).
\end{align}
It is possible to calculate the stabilizable states $\rho_{\mp \mp}$ directly from the left- and right- eigenstates of $\widetilde{H}$ for $\epsilon\neq0$, this is cumbersome due the imaginary part under the square root. In practice it is easier to diagonalize ${\cal W}_0$ directly in the spin basis. Due to normalization the $n_{\text{occ}}$ components of the diagonalized occupied  ${\cal W}_0$- eigenstates are equal to one. Since we are interested in the two occupied eigenstates corresponding to pure states, the normalized $\ew{\sigma_i}$, $i \in \{x,y,z\}$ components have to be real and $\sqrt{\ew{\sigma_x}^2+\ew{\sigma_y}^2+\ew{\sigma_y}^2}=1$. The eigenvalues of these state are $\Lambda_{\mp\mp}=-\frac{1}{2}\Gamma_R\mp\frac{\sqrt{2}}{4}\sqrt{\Gamma_R^2-4\epsilon^2-16T_C^2+\Gamma_W}$ with  $\Gamma_W=\sqrt{(4\epsilon^2+\Gamma_R^2+16T_C^2)^2-64 \Gamma_R^2 T_C^2}$ with       
\begin{align}
\ew{\sigma_x}=&- 
\frac{4 \epsilon^2 +\Gamma_R^2-16T_C^2-\Gamma_W}{16\epsilon T_C}\ew{\sigma_z}, \nonumber \\
\ew{\sigma_y}=& \frac{4\epsilon^2+\Gamma_R^2+16T_C^2-\Gamma_W}{8\Gamma_R T_C}, \nonumber \\
\ew{\sigma_z}=& \mp\frac{\sqrt{\Gamma_R^2-4\epsilon^2-16T_C^2+\Gamma_W}}{\sqrt{2}\Gamma_R}.
\end{align}
In the limit $\epsilon\rightarrow0$ these eigenstates reduce to \eq{smallTc} for $\Gamma_R>4T_C$ and to \eq{bigTc} for $\Gamma_R<4T_C$.
In order to calculate the control operations, these components have to be equal to matching components of $\frac{{\cal J}_L^C}{\Gamma_L} \rho_{00}$:
\begin{align} \label{con-JL}
\ew{\sigma_x}=&\sin(2 \theta) \sin(\theta_C)^2, \nonumber \\
\ew{\sigma_y}=&\sin(\theta)\sin(2\theta_C), \nonumber \\
\ew{\sigma_z}=&\cos(\theta)^2+\cos(2\theta_C)\sin(\theta)^2.
\end{align}
Careful analysis shows:
For $\epsilon>0$ one has $2\Im{(\varepsilon_{\mp})}=\Lambda_{\mp\mp}$, meaning the eigenstate $\rho_{\mp \mp}$ corresponds to the eigenvalue $\Lambda_{\mp\mp}$ of the free Liouvillian. 
While for $\epsilon<0$ we have $2\Im{(\varepsilon_{\mp})}=\Lambda_{\pm\pm}$ and the eigenstate $\rho_{\mp \mp}$ corresponds to the eigenvalue $\Lambda_{\pm\pm}$ of the free Liouvillian.

 

\end{document}